\documentclass[twocolumn,showpacs,preprintnumbers,amsmath,amssymb]{revtex4}

\usepackage{graphicx}
\usepackage{dcolumn}
\usepackage{bm}

\begin{document}

\preprint{cond-mat/xxxxxxxx}

\title{%
Electronic structure and spontaneous internal field 
around non-magnetic impurities 
in spin-triplet chiral $p$-wave superconductors}

\author{Mitsuaki Takigawa}
 \email{takigawa@fcs.coe.nagoya-u.ac.jp}
 \affiliation{%
Department of Computational Science and Engineering, 
21st century COE, Nagoya University, Nagoya 464-8601, Japan
}%
\author{Masanori Ichioka}%
\affiliation{%
Department of Physics, Okayama University, Okayama 700-8530, Japan
}%
\author{Kazuhiko Kuroki}%
\affiliation{%
Department of Applied Physics and Chemistry, 
the University of Electro-Communications, Chofu, Tokyo 182-8585, Japan
}%
\author{Yukio Tanaka}
\affiliation{%
Department of Material Science and Technology, 
Nagoya University, Nagoya 464-8603, Japan
}%

\date{\today}

\begin{abstract}
The electronic structure around an impurity in spin triplet
$p$-wave superconductors is studied by the Bogoliubov-de Gennes theory 
on a tight-binding model, where we have chosen 
$\sin{p_x}{+}{\rm i}\sin{p_y}$-wave or 
$\sin{(p_x+p_y)}{+}{\rm i}\sin{({-}p_x{+}p_y)}$-wave states  
which are considered to be 
candidates for the pairing state in Sr$_{2}$RuO$_{4}$. 
We calculate the spontaneous current
and the local density of states around the impurity
and discuss the difference between the two types of pairing. 
We propose  that it is possible to discriminate the two pairing states  
by studying the  spatial dependence of the 
magnetic field  around a pair of impurities. 
\end{abstract}

\maketitle

The discovery of superconductivity in Sr$_2$RuO$_4$ 
by Maeno {\it et al.} \cite{Maeno} has 
highly activated the field of spin-triplet superconductivity\cite{Sigrist}. 
There are several key  evidences suggesting that 
spin-triplet pairing is realized in Sr$_2$RuO$_4$
\cite{ishida,duffy,nelson,asano,jin,honer,yamash}. 
The presence of 
spontaneous magnetic field suggested from $\mu$SR experiment 
is consistent with a 
chiral superconductivity with broken 
time reversal symmetry (BTRSS) \cite{Luke}. 
Tunneling spectroscopy with zero bias conductance peak \cite{TK}
is also consistent with spin-triplet pairing with BTRSS 
\cite{Mao}. 
\par
Theoretically, several pairing states and/or microscopic mechanisms for 
spin-triplet $p$-wave pairing have been proposed
\cite{Miyake,Kuwabara,Arita,Nomura,Yanase}. 
Although there are three bands in Sr$_{2}$RuO$_{4}$, 
it is possible to consider  the essence of the 
superconducting property only by considering 
quasi two-dimensional $\gamma$ -band. 
For simplicity, here we classify 
the pairings for Sr$_{2}$RuO$_{4}$ into two types,
which are qualitatively consistent with 
experiments of specific heat.\cite{Deguchi} 
The first type is $\sin{p_x}{+}{\rm i}\sin{p_y}$-wave 
proposed by Miyake and Narikiyo (MN), 
where cooper pairs are formed between nearest neighbor sites. 
The second one is $\sin{(p_x{+}p_y)}{+}{\rm i}\sin{({-}p_x{+}p_y)}$-wave state 
proposed by Arita, Onari, Kuroki, and Aoki (AOKA)\cite{Arita}, 
where pairs are formed between next nearest neighbor sites. 
The pairing which has been proposed based on 
third order perturbation theory \cite{Nomura,Yanase} has more higher harmonics 
where the most dominant component is 
$\sin{(p_x{+}p_y)}{+}{\rm i}\sin{({-}p_x{+}p_y)}$-wave pairing. 
In the following, we call  $\sin{p_x}{+}{\rm i}\sin{p_y}$-wave pairing 
and $\sin{(p_x{+}p_y)}{+}{\rm i}\sin{({-}p_x{+}p_y)}$-wave pairing 
as MN-type pairing and AOKA-type, respectively. 
The discrimination between the MN-type and the AOKA-type pairings  
is important because it is strongly  related 
to the study of the superconducting properties  
and the pairing mechanism of Sr$_{2}$RuO$_{4}$. 
One of the remarkable difference of the two pairing states is 
the winding number of the pairing function when we trace its phase along 
the $\gamma$-band Fermi surface \cite{TakigawaP}. 
The winding number is one in the MN-type pairing, but 
three in the AOKA-type pairing. 
It is interesting to propose a new idea to discriminate 
these two pairings having different topological characters. 
\par
The aim of the present paper is to propose that 
experimental observation of the electronic properties and the internal field 
around the impurity can be used to discriminate the 
two pairing states. 
Actually, as shown in recent experiments by 
Lupien {\it et al.},  it is possible to 
observe the local density states in  
Sr$_{2}$RuO$_{4}$ with high accuracy \cite{Lupien}. 
It is shown in the previous studies that 
the electronic structure is modulated 
around the impurity \cite{Salkola,Tsuchiura,Balatsky}. 
For superconductors with BTRSS, 
circular current is induced around impurities \cite{Okuno}
and a spontaneous field appears by a non-magnetic impurity. 
In this paper, we calculate the local density of states (LDOS) 
and spontaneous internal field around the non-magnetic impurity, 
and discuss the difference between 
the two pairings. 
%
\par 
%
To obtain the wave functions and the eigenenergies around the impurity,
we solve the Bogoliubov-de Gennes (BdG) equation on a tight binding model 
which takes into account 
the anisotropy of the Fermi surface and the superconducting gap
structure~\cite{TakigawaP},
\begin{equation}
\sum_i
\left( \begin{array}{cc}
K_{ji} & \Delta_{ij} \\ \Delta_{ij}^\dagger & -K^\ast_{ji}
\end{array} \right)
\left( \begin{array}{c} u_\epsilon({\bf r}_i) \\ v_\epsilon({\bf r}_i)
\end{array}\right)
=E_\epsilon
\left( \begin{array}{c} u_\epsilon({\bf r}_j) \\ v_\epsilon({\bf r}_j)
\end{array}\right) ,
\label{eq:BdG1}
\end{equation}
where
$K_{i,j}=-{t}_{i,j} -\mu \delta_{i,j} $
and $\epsilon$ is an index of the eigenstates.
We set $t_{i,j}=t$, ($-0.4t$) for the transfer
between nearest (second nearest) neighbor sites
in the two dimensional square lattice of ${\rm Ru}$ atoms,
and the chemical potential $\mu$ which depends on the temperature
is tuned so that the density of electron is fixed 
at ${\langle}n{\rangle}=4/3$.
This reproduces the Fermi surface topology of the 
$\gamma$-sheet in ${\rm Sr_2RuO_4}$.
The energy and the temperature are scaled by $t$ throughout this paper.
The self-consistent condition is
\begin{eqnarray} &&
\Delta_{ij}=g_{z,ji} ( \langle a_{j\downarrow} a_{i\uparrow} \rangle
+  \langle a_{j\uparrow} a_{i\downarrow} \rangle  )
\label{eq:SCdz}
\end{eqnarray}
with
$\langle a_{j\downarrow} a_{i\uparrow} \rangle
=\sum_\epsilon v^\ast_{\epsilon}({\bf r}_j)
               u_{\epsilon}({\bf r}_i) f(E_\epsilon)
$ and 
$\langle a_{j\uparrow} a_{i\downarrow} \rangle
=\sum_\epsilon u_{\epsilon}({\bf r}_j)
              v^\ast_{\epsilon}({\bf r}_i)  f(-E_\epsilon),
$
where $f(E)$ is the Fermi distribution function,
and $g_{z,ji}$ is the spin triplet pairing interaction.

The orbital part of the pair potential at each site $i$
can be decomposed into $\sin{p_x}$- and $\sin{p_y}$-components as
\begin{eqnarray} &&
\Delta_{p_x}({\bf r}_i) =
( \Delta_{i,i{+}\hat{x}}
 -\Delta_{i,i{-}\hat{x}} )/2,
\label{eq:dpx}
\\ &&
\Delta_{p_y}({\bf r}_i) =
( \Delta_{i,i{+}\hat{y}}
 -\Delta_{i,i{-}\hat{y}} )/2
\label{eq:dpy}
\end{eqnarray}
as $g_{z,ij}$ is non-zero only for nearest neighbor (NN) site pair 
in the MN-type pairing.
For $\sin{p_x}{\pm}{\rm i}\sin{p_y}$-wave superconductivity,
we define the pair potential as $\Delta_{\pm}({\bf r}_i)
\equiv \Delta_{p_{x}}({\bf r}_i){\pm}{\rm i}\Delta_{p_{y}}({\bf r}_i)$.
When the pair potential is uniform, this BdG formulation is reduced to the
conventional theory for $p$-wave superconductors with the MN-type
pairing functions, which has an anisotropic gap
$(\sin^2{p_x}{+}\sin^2{p_y})^{1/2}$.
In the AOKA-type pairing case, as $g_{z,ij}$
is non-zero only for second nearest site pairs, 
the pair potential $\Delta'_{\pm}({\bf r}_i)$
is defined by $\Delta_{i,i{+}(\hat{x}+\hat{y})}$ and
$\Delta_{i,i{+}({-}\hat{x}{+}\hat{y})}$ instead of 
$\Delta_{i,i{+}\hat{x}}$ and $\Delta_{i,i+\hat{y}}$.
\par 
To investigate the electronic structure around the impurity,
we calculate the LDOS at the $i$-th site as
\begin{eqnarray}
N(E,{\bf r}_i)
=\sum_\epsilon
\{
|u_\epsilon({\bf{r}}_i)|^2\delta(E-E_\epsilon)
+ |v_\epsilon({\bf{r}}_i)|^2\delta(E+E_\epsilon)
\}.
\label{eq:LDOS}
\end{eqnarray}
We evaluate the spontaneous internal field ${\bf H}({\bf r})$ 
through the Maxwell equation:
$\nabla \times {\bf H}= \frac{4\pi}{c} {\bf j}({\bf r})$.
The current ${\bf j}({\bf r})$ is calculated as
\begin{eqnarray}
j_{\hat{e}}({\bf r}_i)
&=&
2 |e| c {\rm Im}\{ {\tilde t}_{i{+}\hat{e},i}
\sum_\sigma \langle a^\dagger_{i{+}{\hat e},\sigma}
a_{i,\sigma} \rangle \} \nonumber \\
&=&
2 |e| c {\rm Im}\{ {\tilde t}_{i{+}{\hat e},i}
\sum_\alpha [ u^\ast_\alpha({\bf r}_{i{+}{\hat e}})u_\alpha({\bf r}_{i})
f(E_\alpha) \nonumber \\
&+& v_\alpha({\bf r}_{i{+}{\hat e}})v^\ast_\alpha({\bf r}_{i})
(1-f(E_\alpha)) ] \}
\label{eq:current}
\end{eqnarray}
for the $\hat{e}$-direction bond ($\hat{e}=\pm\hat{x}$, $\pm\hat{y}$)
at site ${\bf r}_i$.
The spontaneous current $j_{\hat{e}}({\bf r}_i)$
circles around the impurity when time reversal symmetry is broken.
\par 
We consider a system with a square unit cell of $N_r{\times}N_r$ sites 
where an impurity is located at the center of the unit cell.
By introducing the quasi-momentum of the Bloch state,
we obtain the wave function under the periodic boundary condition
whose region covers $N_k{\times}N_k$ unit cells.
We typically consider the case $N_r$=31 and $g_{z,ji}=-t$  
at a low temperature $T=0.01t$.
The chemical potential at the impurity site is taken 
as $\mu_{\rm imp}=-16.0t$.  
The wave function is almost zero at the impurity site. 
\par
\begin{figure}
\vspace{0.5cm} 

\includegraphics[clip,width=75mm]{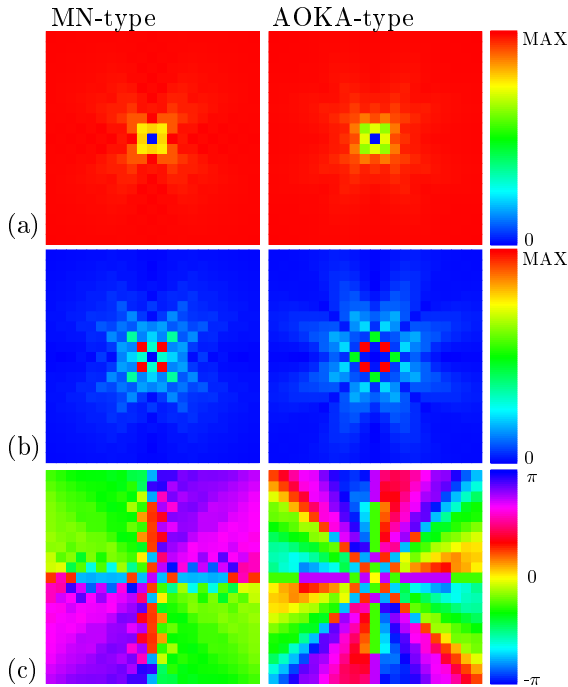}
\caption{\label{fig:1AmpArg}%
(Color) 
Spatial structure of the pairing potential for the MN-type 
(left panels) and the AOKA-type (right panels) pairings.  
We shown a region consisting of $21 \times 21$ sites 
with the impurity located at the center.  
(a) 
Amplitude of the dominant chiral component, 
$|\Delta_{+}({\bf r})|$ or  $|\Delta'_{+}({\bf r})|$. 
(b) Amplitude of the induced opposite chiral component,  
$|\Delta_{-}({\bf r})|$ or $|\Delta'_{-}({\bf r})|$.
(c) Phase structure of the the induced component,  
${\rm Arg}\Delta_{-}({\bf r})$ or 
${\rm Arg}\Delta'_{-}({\bf r})$. 
The phase of the dominant chiral component is almost uniform.
}
\end{figure}

 
First, 
we study the properties around a single impurity, 
and compare the results between the MN-type and the AOKA-type 
pairing cases.
Figure \ref{fig:1AmpArg} shows the pair  potential structure 
around the impurity for MN-type pairing and  for AOKA-type pairing. 
The amplitude of the dominant chiral component 
($\Delta_{+}$ or $\Delta'_{+}$ here) vanishes at the impurity site, 
and recovers within a few site from the impurity, 
as shown in Fig.  \ref{fig:1AmpArg}(a).  
%
On the other hand, the other component with opposite chirality 
($\Delta_{-}$ or $\Delta'_{-}$)  is induced around the impurity
as shown in Fig. \ref{fig:1AmpArg}(b).
%
The phase structure of the induced component, 
shown in Fig. \ref{fig:1AmpArg}(c),   
has a remarkable difference between the two pairing cases,  
reflecting the winding along the Fermi surface. 
%
In the MN case, 
${\rm Arg}\Delta_{-}$ has $+2$-winding at the impurity site,   
and $-1$-winding at four sites located eight sites away from the impurity 
along the vertical and the horizontal directions.  
In the AOKA case, 
${\rm Arg}\Delta'_{-}$ has $-6$-winding at the impurity site, 
and $+1$-windings at four sites located on the diagonal directions 
in addition to $+1$-windings at four sites on the vertical 
and horizontal directions. 
Since the amplitude of the induced component vanishes 
at these winding center sites,    
$|\Delta'_{+}|$ is more suppressed around the impurity with $-6$-winding, 
and the tails of $|\Delta'_{+}|$ have a shape extending toward 
eight directions far from the impurity. 
For the difference of the winding structure of the induced chiral component,  
there appears some differences between the MN-type and the AOKA-type pairings  
in the structure around impurities.

\begin{figure}
\vspace{0.5cm} 

\includegraphics[clip,width=75mm]{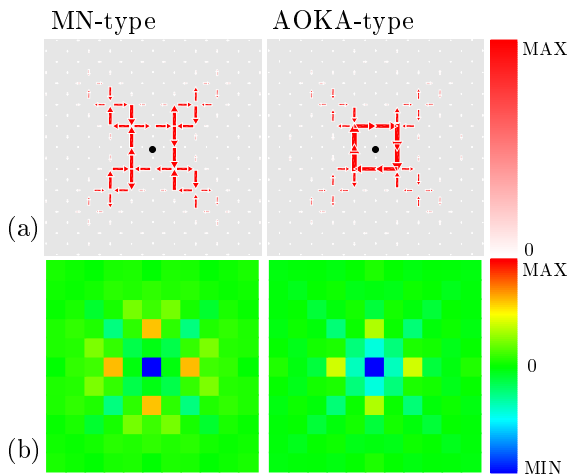}
\caption{\label{fig:1JFldb1}
(Color) 
The spontaneous current $j_e({\bf r})$ (a) and 
the internal magnetic field distribution $H({\bf r})$ (b) 
around a single impurity for the 
MN-type (left panels) and the AOKA-type (right panels) pairings.
We show a region consisting of 
$11 \times 11$ sites with the impurity at the center.
Solid circles in (a) indicate the impurity site. 
}
\end{figure}

In Fig. \ref{fig:1JFldb1}(a),
we plot the spontaneous current $j_{\hat{e}}({\bf r})$ around the impurity. 
The current circles around the impurity, 
and the staggered current spreads toward the diagonal directions.
The current in the diagonal direction is larger in the MN case. 
We show the internal magnetic field in Fig. \ref{fig:1JFldb1}(b). 
The spontaneous field at the impurity site is oriented to the $-z$ direction. 
This induced field is larger in the AOKA case, since the the spontaneous 
current around the impurity is larger [Fig. \ref{fig:1JFldb1}(a)] due to 
the $-6$-winding of $\Delta'_{-}$ at the impurity 
[Fig. \ref{fig:1AmpArg}(c)]. 
To compensate the spontaneous field at the impurity site, 
the spontaneous field is oriented toward 
the $+z$ direction near the impurity 
along the horizontal and vertical directions.


\begin{figure}
\vspace{0.5cm} 

\includegraphics[clip,width=75mm]{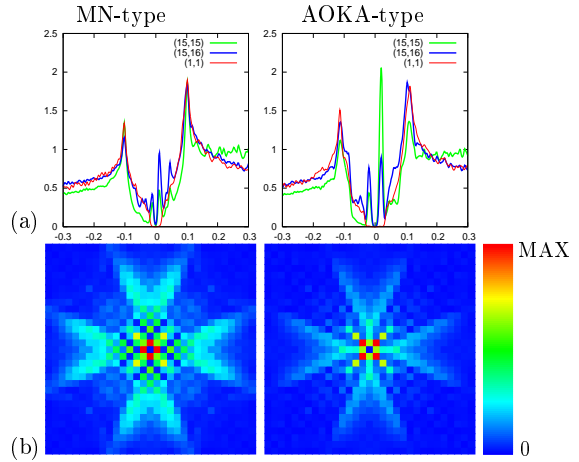}
\caption{\label{fig:1Ldos-ER} 
(Color) 
(a) Spectrum of the LDOS $N(E,{\bf r})$ at sites  (15,15), (15,16) and (1,1).  
(b) Density plot of the LDOS $N(E{\sim}0,{\bf r})$ at the peak energy of 
the spectrum near $E\sim 0$ within a unit cell of $31 \times 31$ sites. 
The left and right panels are, respectively, for the MN-type 
and the AOKA-type pairings.  
}
\end{figure}

Figure \ref{fig:1Ldos-ER}(a) shows the LDOS $N(E,{\bf r}_i)$ 
around the impurity. 
The impurity is located at site (16,16). 
We plot the $N(E,{\bf r}_i)$ at the nearest site ${\bf r}_i=(15,16)$,
at the second nearest site (15,15), and at the farthest site (1,1).
The energy spectrum of the uniform state with $p$-wave anisotropic gap is 
reproduced at (1,1). 
Splitted LDOS peaks appear within the gap at (15,15) and (15,16) 
near the impurity. 
The side peak in the positive energy range is larger than that in the 
negative range for both pairings. 
The peak at (15,15) is larger (smaller) than that at (15,16)
in AOKA-type (MN-type) pairing. 
The LDOS at this peak energy is shown 
in Fig. \ref{fig:1Ldos-ER}(b). 
The tail structure far from the impurity is similar,
but the LDOS at nearest and second nearest sites 
is different between MN-type and AOKA-type pairings.
In the case of AOKA-type (MN-type) pairing between diagonal (vertical) 
site quasiparticles, 
the low energy state is likely to appear in the diagonal (vertical) site 
next to the impurity site. 
The anisotropic gap structure on the Fermi surface 
also contributes to the difference of the LDOS.
In the AOKA-type pairing, the gap amplitude has local minimum 
at [110] directions on the Fermi surface 
in addition to the minimum at [100] directions,  
as shown in Fig. 1 of ref. \cite{TakigawaP}.

\begin{figure}[htb]
\vspace{0.5cm} 

\includegraphics[clip,width=75mm]{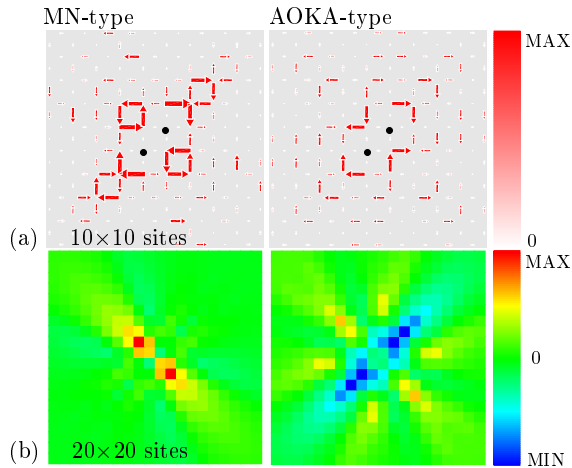}
\caption{\label{fig:ArgJHb1}%
(Color) 
The spontaneous current $j_e({\bf r})$ (a) and 
the internal magnetic field $H({\bf r})$ (b) 
for the MN-type (left panels) and the AOKA-type (right panels) pairings     
when a pair of impurities is situated in the diagonal direction.    
e show a region of $10 \times 10$ sites for (a) 
and a region of $20 \times 20$ sites for (b) around the impurities. 
Solid circles in (a) indicate the impurity sites. 
}
\end{figure}

As we have seen in Fig. \ref{fig:1JFldb1}(b), 
a drastic qualitative difference is not seen in the internal field 
between the two pairings in the case of single impurity.
However, in multiple impurity case, there are some cases producing  
qualitative differences between the two pairing cases 
in the spontaneous field distribution around the impurities  
due to the interference of the phase of the 
induced opposite chiral component 
(i.e. $\Delta_{-}({\bf r})$ or $\Delta'_{-}({\bf r})$ 
in Fig. \ref{fig:1AmpArg}). 
As an example of this effect, we report the case when  
two impurities are located at next nearest neighboring sites,
i.e., sites (15,15) and (16,16) in our calculation. 
As in the single impurity case,  the zero energy LDOS 
is large at the nearest site of the impurities for MN-type, 
and at the next nearest site for AOKA-type.
As shown in Fig. \ref{fig:ArgJHb1}, the spontaneous current and the field 
distribution is not a simple summation of the distribution of the single 
impurity case, 
since the strong interference is at work between the two impurities.  
It is remarkable that 
for MN-type pairing, 
the field intensity is strong in the direction perpendicular  
to the direction in which the impurities are aligned. 
The field orientation is toward the $+z$ direction in this case.  
On the other hand, in AOKA-type pairing, 
the spontaneous field, oriented toward the $-z$ direction,  
has strong intensity in the direction parallel to the impurity 
alignment. 
This is a good example of phenomena in which 
a difference in the phase windings of the pairing function 
on the Fermi surface results in experimentally observable quantities.
\par 
In summary, 
we have calculated, on the basis of BdG equation, 
the spontaneous field and the electronic structure
around the impurities in spin triplet chiral $p$-wave superconductor,  
where we compare the results between 
the MN-type pairing $\sin{p_x}{+}{\rm i}\sin{p_y}$ 
and the AOKA-type pairing $\sin{(p_x{+}p_y)}{+}{\rm i}\sin{({-}p_x{+}p_y)}$. 

We have shown 
that differences between the two pairings appear in the LDOS structure  
at the sites nearest and next nearest to the impurity site.
Namely, the zero energy LDOS at the nearest site is found to be 
larger (smaller) than 
the LDOS at the second nearest site for MN-type (AOKA-type) pairing.
When a pair of impurities is situated in the diagonal direction, 
the spontaneous magnetic field distribution 
around the impurities is quite different between the two pairings. 
The spontaneous field has strong intensity along the line 
perpendicular (parallel) to the direction in which the two impurities are 
aligned in the MN-type (AOKA-type) pairing. 
We expect that the properties around the impurities 
shown here provide information for 
distinguishing the pairing state of Sr$_2$RuO$_4$.
\par 
%
The authors thank Y. Maeno, K. Ishida, H. Kontani, and Y. Yanase 
for valuable discussions. 
They also thank H. Aoki and R. Arita for stimulating discussions.
This work was supported by a Grant-in-Aid
for the 21st Century COE "Frontiers of Computational Science".

\end{document}